\documentclass[oldversion]{aa}  
\usepackage{graphicx}
\usepackage{txfonts}
\begin{document}
   \title{Star cluster versus field star formation in the nucleus of the prototype starburst galaxy M82}
   \titlerunning{Nuclear star and cluster formation in M82}

   \author{S. Barker$^{1,2}$, R. de Grijs$^{1,3}$
          \and
          M. Cervi\~no$^4$
          }
   \authorrunning{S. Barker, R. de Grijs and M. Cervi\~no}

   \offprints{R. de Grijs}

   \institute{$^1$ Department of Physics \& Astronomy, The University of
   Sheffield, Hicks Building, Hounsfield Road, Sheffield S3 7RH, UK\\
   $^2$ Isaac Newton Group of Telescopes, Apartado de correos 321,
   E-38700 Santa Cruz de la Palma, Canary Islands, Spain\\
   $^3$ National Astronomical Observatories, Chinese Academy
   of Sciences, 20A Datun Road, Chaoyang District, Beijing 100012,
   China\\
   $^4$ Instituto de Astrof{\'\i}sica de Andaluc{\'\i}a (CSIC), Camino
   Bajo de Hu\'etor, 50, Granada, E-18008, Spain\\
   \email{pha04sb@sheffield.ac.uk, R.deGrijs@sheffield.ac.uk, mcs@iaa.es}}

   \date{Received / accepted}

   \abstract{We analyse high-resolution {\sl Hubble Space
Telescope}/Advanced Camera for Surveys imaging of the nuclear
starburst region of M82, obtained as part of the Hubble Heritage
mosaic made of this galaxy, in four filters (Johnson-Cousins
equivalent $B, V$, and $I$ broad bands, and an H$\alpha$ narrow-band
filter), as well as subsequently acquired $U$-band images. We find a
complex system of $\sim$150 star clusters in the inner few 100 pc of
the galaxy. We do not find any conclusive evidence of a
cluster-formation epoch associated with the most recent starburst
event, believed to have occurred about 4--6 Myr ago. This apparent
evidence of decoupling between cluster and field-star formation is
consistent with the view that star cluster formation requires special
conditions. However, we strongly caution, and provide compelling
evidence, that the `standard' simple stellar population analysis
method we have used significantly underestimates the true
uncertainties in the derived ages due to stochasticity in the stellar
initial mass function and the corresponding sampling effects.}


   \maketitle

\section{Introduction}

Supermassive star clusters or `super star clusters' (SSCs) are highly
luminous and massive, yet compact star clusters. They result from the
most intense star-forming episodes that are believed to occur at least
once in the lifetime of nearby starburst galaxies, and most probably
also in merging galaxies at high redshift (Smith et al. 2006). Super
star clusters have been observed using the {\sl Hubble Space Telescope
(HST)} in interacting, amorphous, dwarf, and starburst galaxies,
(e.g., Arp \& Sandage 1985; Melnick, Moles \& Terlevich 1985; Holtzman
et al. 1992; Meurer et al. 1992; Whitmore et al. 1993, 1999; Hunter et
al. 1994, 2000; O'Connell, Gallagher \& Hunter 1994; Ho \& Filippenko
1996; Conti, Leitherer \& Vacca 1996; Watson et al 1996; Ho 1997;
Carlson et al. 1998; de Grijs, O'Connell \& Gallagher 2001; de Grijs
et al. 2003a,b; and references therein) and are thought to be
contemporary analogues of young globular clusters. For this reason,
the study of SSCs can lead to an understanding of the formation,
evolution, and destruction of globular clusters (see the review by de
Grijs \& Parmentier 2007) as well as of the processes of star
formation in extreme environments.

A current issue of strong contention relates to the stellar initial
mass function (IMF) in starburst environments, and particularly
whether it is `abnormal' due to preferential production of higher-mass
stars (we will touch on this issue below, where we explore the effects
of stochasticity in the IMF). Furthermore, SSCs are involved in the
activation and feeding of supergalactic winds (e.g., Tenorio-Tagle,
Silich \& Mu\~noz-Tu\~n\'on 2003; Westmoquette et al. 2007a,b). In
addition, SSCs can be age-dated individually using either
spectroscopic or multi-passband imaging observations, and as such they
can be used as powerful tracers of the starburst history across a
given galaxy.

The target galaxy discussed in this paper, M82, is the prototype
nearby starburst galaxy. The starburst is five times as luminous as
the entire Milky Way, and it has one hundred times the luminosity of
the centre of our Galaxy (O'Connell et al. 1995; and references
therein). At a distance of 3.9 Mpc (Sakai \& Madore 1999), many of
M82's individual clusters can easily be resolved by the {\sl
HST}. O'Connell et al. (1995) performed an optical imaging study of
the central region of M82 using the {\sl HST}'s Wide Field / Planetary
Camera (WF/PC), and unveiled over one hundred candidate SSCs (in their
nomenclature) within the visible starburst region, with a mean $M_V
\sim -11.6$ mag (based on an adopted distance of 3.6 Mpc). At the
present time, this is brighter than any globular cluster in the Local
Group (e.g., Ma et al. 2006; and references therein). A more recent
study by Melo et al. (2005) used {\sl HST}/Wide Field and Planetary
Camera-2 (WFPC2) observations to reveal 197 young massive clusters in
the starburst core (with a mean mass close to $2 \times 10^5$
M$_\odot$, largely independent on the method used to derive these
masses), confirming this region as a most energetic and high-density
environment (see also Mayya et al. 2008).

Unfortunately, studies of the core are inhibited by M82's almost
edge-on inclination ($i \sim 80^\circ$; Lynds \& Sandage 1963; McKeith
et al. 1995) and the great deal of dust within the galactic disk (see
Mayya et al. 2008 and references therein). Although the dust provides
the raw material for the clusters' formation, it also obscures and
reradiates most of the light from the starburst in the infrared (Keto
et al. 2005). However, despite all the bright sources associated with
the active starburst core suffering from such heavy extinction (in the
range $A_V \sim 5$--25 mag; e.g., Telesco et al. 1991; McLeod et
al. 1993; Satyapal et al. 1995; Mayya et al. 2008, and references
therin), many individual clusters remain bright enough to obtain good
photometry and spectroscopy for (see, for example, Smith et al. 2006;
McCrady \& Graham 2007).

The starburst galaxy M82 has recently experienced at least one tidal
encounter with its large spiral neighbour galaxy, M81, resulting in a
large amount of gas being channelled into the core of the galaxy over
the last 200 Myr (e.g., Yun, Ho \& Lo 1994). The most major recent
encounter is believed to have taken place roughly $(2 - 5) \times
10^8$ yr ago (Brouillet et al. 1991; Yun et al. 1994; see also Smith
et al. 2007), causing a concentrated starburst and an associated
pronounced peak in the cluster age distribution (e.g., de Grijs et
al. 2001, 2003c; Smith et al. 2007). This starburst continued for up
to $\sim 50$ Myr at a rate of $\sim 10$ M$_\odot$ yr$^{-1}$ (de Grijs
et al. 2001). Two subsequent starbursts ensued, the most recent of
which (roughly 4--6 Myr ago; F\"orster Schreiber et al. 2003,
hereafter FS03; see also Rieke et al. 1993) may have formed at least
some of the core clusters (Smith et al. 2006) -- both SSCs and their
less massive counterparts (see Section 3.6).

The active starburst region in the core of M82 has a diameter of 500
pc and is defined optically by the high surface brightness regions, or
`clumps', denoted A, C, D, and E by O'Connell \& Mangano (1978). These
regions correspond to known sources at X-ray, infrared, and radio
wavelengths, and as such they are believed to be the least obscured
starburst clusters along the line of sight (O'Connell et
al. 1995). The galaxy's signature bi-polar outflow or `superwind'
appears to be concentrated on regions A and C (Shopbell \&
Bland-Hawthorn 1998; Smith et al. 2006; Silich et al. 2007), and is
driven by the energy deposited by supernovae, at a rate of $\sim 0.1$
supernova yr$^{-1}$ (e.g., Lynds \& Sandage 1963; Rieke et al. 1980;
Fabbiano \& Trinchieri 1984; Watson, Stranger \& Griffiths 1984;
McCarthy, van Breugel \& Heckman 1987; Strickland, Ponman \& Stevens
1997; Shopbell \& Bland-Hawthorn 1998; see also de Grijs et al. 2000
for a review)

This paper is organised as follows. In Section 2 we give an overview
of the observations, source selection, and image processing techniques
applied to our sample of star clusters in the core of M82. Section 3
describes the methods and results of the age determinations, and the
complications due to stochasticity in the IMF, and Section 4 provides
a summary and discussion of these results, where we place our star
cluster results in the context of the star-formation history of the
galaxy.

\section{Source Selection and Photometry}

\subsection{Observations}

During March 2006, a large four-filter six-point mosaic data set of
M82 was obtained as part of the Hubble Heritage Project ({\sl HST}
proposal GO-10776) using the Advanced Camera for Surveys/Wide Field
Channel (ACS/WFC; pixel size $\sim 0.05$ arcsec). Exposure times were
1600, 1360, 1360, and 3320 s for observations in the F435W (equivalent
to Johnson $B$), F555W ($\sim V$), F814W ($\sim I$), and F658N
(H$\alpha$) filters, respectively. For each of these filters, four
exposures were taken at six marginally overlapping pointings, or
`tiles'. The four exposures within each tile were dithered to improve
the rejection of detector artefacts and cosmic rays, and to span the
interchip gap of the ACS/WFC. The final mosaic of the $3 \times 2$
ACS/WFC fields was pipeline-processed by the {\sl HST} data archive's
standard reduction routines; for full details of the observing
programme, image processing and calibration, see Mutchler et
al. (2007).

Whereas O'Connell et al. (1995) had access to two broad-band filters
only (F555W and F785LP, the latter corresponding to a broad $\sim
R$-band filter), the present study uses these four-filter
high-resolution ACS Hubble Heritage observations of M82 as well as
subsequently acquired F330W ($\sim U$-band) data, observed as part of
proposal GO-10609 (PI Vacca), and obtained from the {\sl HST} data
archive. The F330W observations centred on M82 A and C were obtained
on UT December 7 and 8, 2006, respectively, with the ACS/High
Resolution Camera (ACS/HRC; pixel size $\sim 0.028 \times 0.025$
arcsec). The respective exposure times were 4738 and 3896 s, with the
ACS/HRC pointed under a position angle of 90.08 and 90.06$^\circ$
(East with respect to North), respectively. The two final
pipeline-processed and flux-calibrated images cover adjacent fields,
largely coincident with the central starburst area selected for
further study in the current paper (see Section 2.2).

It should also be noted that O'Connell et al. (1995) used pre-COSTAR
data, acquired before the first {\sl HST} servicing mission. The
increased volume and quality of data now available to us provides more
evidence with which to distinguish clusters from point sources (i.e.,
stars), and in principle allows for better constrained age estimates,
making the data set ideally matched for the first comprehensive
comparison of the various star-formation modes in one of the most
violent environments in the local Universe.

\subsection{Basic Image Processing}

\begin{figure}
\includegraphics[width=\columnwidth]{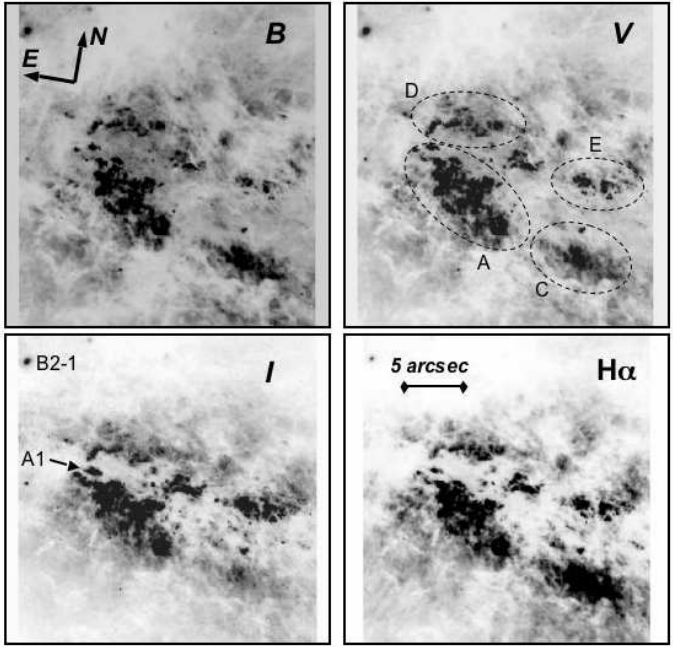}
\caption{The starburst core region of M82 used for the main analysis
in this paper. The four panels show the region's appearance as a
function of wavelength ($B$: F435W; $V$: F555W; $I$: F814W; H$\alpha$:
F658N). Each panel is 29.55 arcsec on a side; a scale bar of 5 arcsec
length (corresponding to $\sim 95$ pc at the distance of M82) is shown
in the bottom right-hand panel. In the top right-hand panel, we show
the outlines of regions A, C, D, and E, defined by O'Connell \& Mangano
(1978), superposed on the $V$-band image. The bottom left-hand panel
includes the positions of clusters A1 and B2-1 (see Smith et
al. 2006). The individual panels are displayed at optimum dynamic
range, in order to emphasise the low-level structure.}
\label{m82core.fig}
\end{figure}

The final drizzled $B, V, I$, and H$\alpha$ images were first aligned
using the {\sc iraf/stsdas} tasks\footnote{The Image Reduction and
Analysis Facility ({\sc iraf}) is distributed by the National Optical
Astronomy Observatories, which is operated by the Association of
Universities for Research in Astronomy, Inc., under cooperative
agreement with the U.S. National Science Foundation. {\sc stsdas}, the
Space Telescope Science Data Analysis System, contains tasks
complementary to the existing {\sc iraf} tasks. We used Version 3.5
(March 2006) for the data reduction performed in this paper.}  {\sc
imalign} and {\sc rotate}, using a selection of conspicuous sources
common to each frame as guides. They were then cropped to a common
size of $541 \times 591$ pixels ($27.05 \times 29.55$ arcsec) in order
to include only the very core of the galaxy, the region of particular
interest in this paper. The $U$-band images were rotated, aligned and
scaled to the Hubble Heritage data set using a similar selection of
point-like sources common to both data sets, where available. This
ensured that the astrometric solutions of all images were calibrated
to a common frame of reference.

\begin{figure}
\includegraphics[width=9cm]{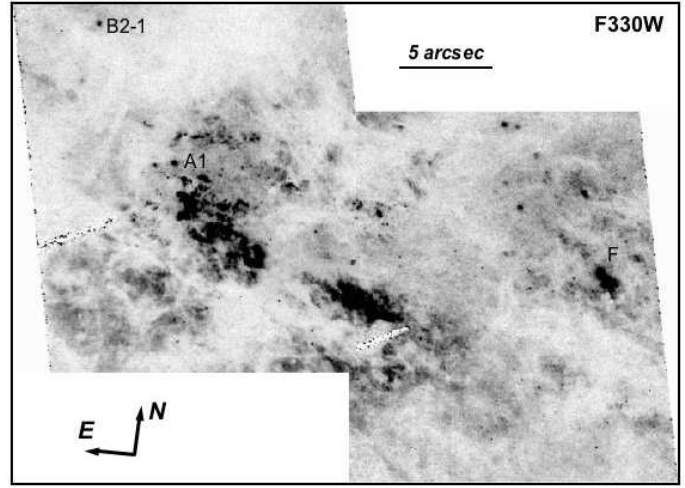}
\caption{Complementary F330W observations of the M82 A (east) and M82
C (west) pointings obtained with the ACS/HRC. The position angle is
identical to that of the four images shown in Fig.  \ref{m82core.fig};
clusters A1, B2-1, and F (e.g., Gallagher \& Smith 1999) are indicated
for orientation. The mosaic is displayed at optimum dynamic range, in
order to emphasise the low-level structure. The linear features on the
left-hand side of each of the two mosaicked pointings, seen under an
angle of about 30$^\circ$, are instrumental artefacts due to the HRC's
coronographic finger. The occulting finger is not retractable and is
therefore present in every ACS/HRC exposure.}
\label{m82uband.fig}
\end{figure}

The selected region was chosen to specifically encompass the areas A,
C, D, and E (O'Connell \& Mangano 1978), as well as cluster B2-1 (de
Grijs et al. 2001; Smith et al. 2006) for reasons of both image
alignment and for photometric consistency checks. The starburst region
studied in detail in this paper is shown in Fig. \ref{m82core.fig}. In
Fig. \ref{m82uband.fig} we show the complementary, mosaicked F330W
ACS/HRC observations, rotated to the same orientation as the fields in
Fig. \ref{m82core.fig}.

At first glance, the images in the individual passbands exhibit a
number of striking similarities as well as differences. It is clear
that the M82 core is undergoing vigorous star formation, as indicated
by the strong H$\alpha$ and F330W-band emission. A first impression of
the various named regions indicates that region C may be the youngest
(since it does not exhibit strong emission in the $I$-band filter),
while region E might be somewhat older or most affected by extinction
(given that it is progressively easier to identify as a function of
increasing wavelength). We will quantify these first impressions in
Section 3.

\subsection{Source Selection}

\begin{figure}
\includegraphics[width=\columnwidth]{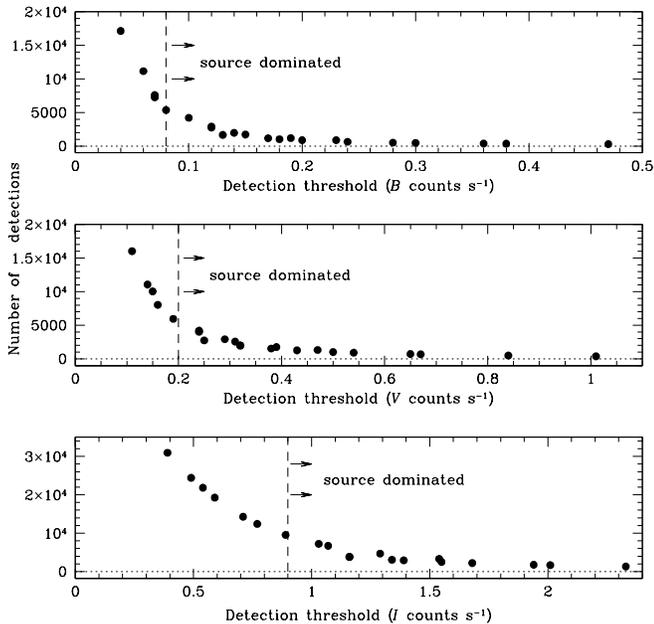}
\caption{Number of objects detected as a function of threshold level
for the $B, V$, and $I$ images. The vertical dashed lines show the
thresholds adopted here for source-dominated (as opposed to
noise-dominated) ranges. See the text for details (Section 2.3).}
\label{threshold.fig}
\end{figure}

The standard deviations of the number of counts of seemingly empty
sections of all images were established to ascertain a `sky'
background count for each filter. Multiples of this background count
were then used as thresholds above which the number of sources in each
filter was calculated, using the {\sc idl}\footnote{The Interactive
Data Language {\sc (idl)} is licensed by Research Systems Inc., of
Boulder, CO, USA.} {\sc find} task. Figure \ref{threshold.fig} shows
that the most suitable, conservatively chosen thresholds for source
inclusion were 0.08, 0.20, and 0.90 counts s$^{-1}$, for the $B, V$,
and $I$ bands, respectively. (In view of the more stringent
requirements applied in the next steps in the construction of our
final sample, the exact values of these thresholds are unimportant, as
long as they are chosen such that no real objects are omitted {\it a
priori} at this stage.) For all three passbands, initially the number
of detections decreases rapidly with increasing threshold value. This
is an indication that our `source' detections are noise
dominated. Where the rapid decline slows down to a more moderate rate,
our detections become dominated by `real' objects (either stars,
clusters, or real intensity variations in the background field). In
Fig.  \ref{threshold.fig} we have indicated the approximate count
rates of these transitions by the vertical dashed lines; in the
remainder of this paper we will only consider the objects in the
`source-dominated' domain, as indicated in the three panels.

We point out that the curve delineated by the data points is not as
smooth as perhaps expected at face value. This is due to our use of a
number of different ranges of relatively low background flux in which
we determined the standard deviations of the counts. We subsequently
ran our {\sc find} routine based on detection thresholds in multiples
of the various standard deviations. The deviations from continuous
smooth curves are therefore a good indication of the intrinsic
variations in seemingly empty field regions at low flux levels.

We subsequently employed a cross-identification procedure to determine
how many sources were visible and coincident with intensity peaks
within 1.4 pixels of each other in all three filters (i.e., allowing
for 1-pixel mismatches in both spatial directions), yielding over 900
potential clusters. Sources found to be present in all filters were
then confirmed individually by eye, owing to the lack of a more robust
verification procedure in the presence of the highly variable
background and significantly variable extinction across our field of
view (cf. de Grijs et al. 2001; Smith et al. 2007). This verification
process was performed independently by two of us, and the small number
of disagreements (less than a few tens of objects) were discussed and
revisited in detail in order to construct the most robust and
impartial cluster sample possible.

\begin{figure}
\includegraphics[angle=-90,width=\columnwidth]{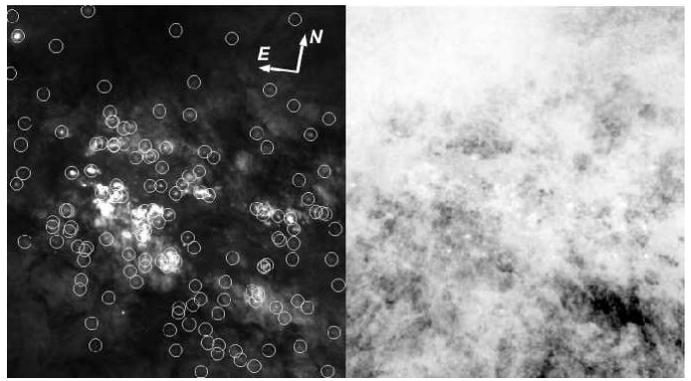}
\caption{{\it (left)} Distribution of our final, cross-matched cluster
sample across the face of the M82 starburst core, superposed on the
$V$-band image. {\it (right)} H$\alpha$ excess distribution. The
darkest shading corresponds to the most significant H$\alpha$ excess;
the lightest shading indicates no (or an insignificant) excess. The
orientation and scale of the images are as in Fig. \ref{m82core.fig}.}
\label{sources.fig}
\end{figure}

In our next step we used the `Tiny Tim' software package (Krist \&
Hook 1997) to generate {\sl HST} point-spread functions (PSFs). We
determined the best-fitting Gaussian profile for each candidate
cluster, as well as for the Tiny Tim PSFs. The width of the latter was
found to be best represented by a Gaussian profile of $\sigma_{\rm G}
= 1.38$ pixels, in the $V$ band. Any candidate cluster with
$\sigma_{\rm G}$ (significantly) below this value was considered most
likely to be a star (or an artefact either of the CCD or due to cosmic
rays) and was consequently discarded. Artificial clusters were then
created using the {\sc BAOlab} package (Larsen 1999), using {\sl
HST}/ACS PSFs, and their $\sigma_{\rm G}$ values were compared with
those determined by our {\sc idl} routine. This provided an additional
constraint with which to distinguish clusters from stars, leaving 208
clusters for further analysis. We note that although realistic cluster
luminosity profiles may well deviate from the simple Gaussian profile
adopted here, its consistent and systematic application to our
observations allows us to robustly differentiate between objects of
different sizes, irrespective of their true profiles, provided that
the profiles of the individual objects do not differ too much from
object to object. In Fig. \ref{sources.fig} we display the
distribution of our final cluster sample across the face of the M82
starburst core in the $V$ band. This final sample was used for further
analysis in {\it all} of the $U, B, V, I$, and H$\alpha$ images. 

The right-hand panel of Fig. \ref{sources.fig} shows the distribution
of the H$\alpha$ excess across the region. To produce this panel, we
constructed the interpolated H$\alpha$ continuum from a combination of
the $V$ and $I$-band images, scaled by their filter widths and
exposure times. The image shows the ratio of the H$\alpha$ image and
its continuum. The darkest shading corresponds to the most significant
H$\alpha$ excess; the lightest shading to no (or a negligible)
excess. We will discuss this image in Section \ref{ages.sect}.

\subsection{Photometry}

Two photometry tasks were used, in {\sc idl}, with the aim of robustly
determining the magnitudes of as many clusters as possible. The {\sc
apphot} task was used with source radii ranging from 1--10 pixels and
a standard sky annulus with radii from 10 to 15 pixels, corresponding
to a linear scale of 10 to 14 pc at the distance of M82. In addition,
our custom-written task {\sc photom} utilises radii individualised for
each cluster (based on visual inspection). The resulting
`instrumental' magnitudes (in the $U, B, V$, and $I$ bands) of the 152
clusters for which our tasks could robustly obtain photometry were
adjusted for their zero-point offsets (determined from the headers of
the archival {\sl HST} observations).

In Anders, Gieles, \& de Grijs (2006) we carefully considered the
implications and remedies of taking `sky' annuli too close to a given
cluster, in the sense that we would oversubtract the sky level because
of the inclusion of flux from the cluster's outer profile. We
generated an extensive and systematic grid of aperture corrections as
a function of filter, radius of the sky annulus, and cluster
profile. Using these improved aperture corrections, we corrected our
cluster aperture photometry based on a fixed source aperture for the
systematic effects introduced by our reduction procedure. Upon close
examination of our aperture choice, we concluded that a 3-pixel
(radius) source aperture resulted in the smallest photometric scatter
for the cluster sample as a whole. We will adopt this source aperture
below. The resulting average (aperture-corrected) magnitudes of the
clusters were 20.0, 19.5, and 19.3 mag in the $B, V$, and $I$ bands,
respectively, with our cluster photometry spanning in excess of four
magnitudes in all filters (5 mag in the $I$ band).

Although the data reduction procedure employed does not allow us to
perform a proper assessment of the observational completeness limits,
the luminosity functions in the $B, V$, and $I$ bands indicate that
our detection limits are at $B \simeq 23.5, V \simeq 23$, and $I
\simeq 24$ mag. As also shown in Fig. \ref{threshold.fig}, where we
showed the number of detected objects as a function of our detection
threshold, our combined $BVI$ photometry is limited by the $B$ and
$V$-band observations.

\begin{figure}
\includegraphics[width=\columnwidth]{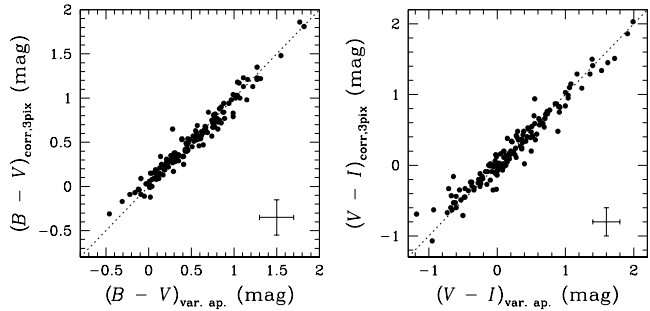}
\vspace{-4.7cm}
\caption{Comparison of our $(B-V)$ and $(V-I)$ colours obtained from
variable aperture photometry according to the actual cluster sizes,
with fixed 3-pixel (radius) aperture photometry and aperture
corrections from Anders et al. (2006). For reasons of clarity we have
not added error bars to the data points; a representative set of error
bars is shown in the bottom right-hand corner of each panel. The
dotted lines of equality are shown to guide the eye.}
\label{colours.fig}
\end{figure}

In Section 3 we will analyse our cluster photometry based on the
cluster colours. In essence, we will use a poor man's approach to
cluster spectroscopy by using the broad-band spectral energy
distributions (SEDs). Since a small change in colour could lead to a
large change in the derived age, it is important to make sure that the
application of the Anders et al. (2006) aperture corrections has not
introduced unwanted systematic effects and/or offsets. Therefore, in
Fig. \ref{colours.fig} we compare the cluster colours derived from our
variable-aperture photometry (obtained with the {\sc photom} task) to
those from the standard apertures (based on {\sc apphot}) combined
with the Anders et al. (2006) aperture corrections. Considering the
representative uncertainties in the derived colours (shown in the
bottom right-hand corners of both panels), both panels of
Fig. \ref{colours.fig} show excellent agreement between the two data
sets. We are therefore confident that these aperture corrections will
not introduce systematic offsets in the age determinations in Section
3.

\section{Cluster ages}

In this section we use the broad-band photometry derived in Section 2
to obtain estimates (or otherwise) of the star cluster ages in the M82
starburst core. For optimum results, we do not only require a
sufficiently high age resolution in our models, but also take into
account the effects on the integrated luminosities of statistically
sampling the stellar IMF (e.g., Cervi\~no, Luridiana, \& Castander
2000; Cervi\~no et al. 2002; Cervi\~no \& Luridiana 2004, 2006). We
will first discuss issues related to statistically sampling the IMF
(Sections 3.1--3.4), and then apply the `standard' analysis,
essentially ignoring these issues, in Sections 3.5 and 3.6. This will
highlight the differences caused by stochasticity in the IMF of
incompletely sampled clusters.

\subsection{Spectral synthesis models and isochrones}
\label{models.sect}

In order to achieve these aims, we have made use of the {\sc cmd2.0}
tool\footnote{available from {\tt http://stev.oapd.inaf.it/cmd}.},
which provides the isochrones and the mean integrated magnitudes for a
number of photometric systems, including {\sl HST}/ACS data in the ST
system used here. We calculated the integrated magnitudes for simple
stellar populations (SSPs) based on a Kroupa (2001) IMF (corrected for
binaries), covering a mass range from 0.15 to 120 M$_\odot$, from the
solar metallicity ($Z=0.019$) isochrones of Girardi et al. (2002) and
Marigo et al. (2008) for stars more massive than 7 M$_\odot$, and the
Bertelli et al. (1994) isochrones for low-mass stars. Near-solar
metallicity should be a reasonable match to the young objects in M82
(e.g., Gallagher \& Smith 1999; see also Smith et
al. 2006). Fritze-von Alvensleben \& Gerhard (1994) also find, from
chemical evolution models, that young clusters should have $Z \sim
0.3$--1.0 Z$_\odot$. Magnitudes and other quantities are based on
Kurucz (1992) atmosphere models, except for cool stars (see Girardi et
al. 2000 for more details).

Our set of SSP models and isochrones covers an age range from 1 Myr to
10 Gyr. We note, as a caveat, that the models do not include nebular
emission, which may systematically affect the results at the youngest
ages (see also Anders \& Fritze-von Alvensleben 2003). We will return
to this issue below (Section 3.7), where we discuss the results from
the H$\alpha$ observations in detail.

Finally, we adopted the Galactic extinction law of Rieke \& Lebofsky
(1985; conveniently tabulated by Jansen et al. 1994) to correct our
cluster photometry for foreground extinction in M82. Referring to
fig. 1 of de Grijs et al. (2005), we note that the choice of
extinction law in the optical wavelength range is relatively
unimportant in the context of extinction corrections of broad-band
photometry. Galactic foreground extinction was estimated based on
Schlegel et al. (1998).

\subsection{The `lowest-luminosity limit' test}

Estimates of physical parameters (such as mass or age) based on
synthesis models assume a certain proportionality between the stars at
different evolutionary phases for a given age. Of course, the use of
the mean integrated luminosity provided by the SSP models only makes
sense when this proportionality is robustly met. This implies that
there must be sufficient numbers of stars in any and all evolutionary
phases or, equivalently, that the IMF is well sampled in all relevant
mass intervals (see Cervi\~no \& Valls-Gabaud 2008 for a more detailed
discussion).

The simplest test to identify whether the IMF is appropriately sampled
is to use the `lowest-luminosity limit' (LLL) method (Cervi\~no \&
Luridiana 2004). The LLL method states that the IMF is not
sufficiently well sampled if the integrated luminosity of a cluster is
lower than the luminosity of the most luminous star included in the
model for the relevant age (see Pessev et al. 2008 for an
application). This implies that clusters fainter than this limit
cannot be analysed using standard procedures, including $\chi^2$
minimisation of the observed values with respect to the mean SSP
models (but see Sections 3.5 and 3.6). Below the LLL, the
proportionality assumed in the mean SSP value is not met, and the
cluster colours do not reflect the cluster age of the entire
population, but -- instead -- the colour combination of a given set of
individual (luminous) stars. Consequently, in this case cluster ages
and masses cannot be obtained self-consistently (although this caveat
is largely ignored in most studies using SED fits to obtain cluster
ages).

\begin{figure}
\includegraphics[width=\columnwidth]{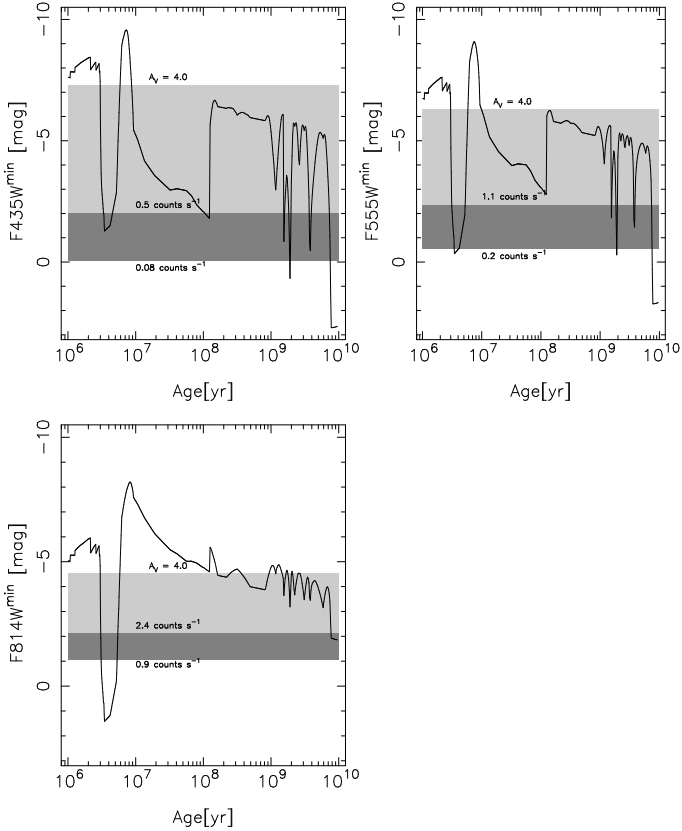}
\caption[]{Lowest-luminosity limit for the ACS/WFC filters used here.
The dark-grey area corresponds to the `source-dominated' clusters
(Fig. 3), assuming a distance to M82 of $D= 3.9$ Mpc. The light-grey
area shows the cluster's maximum absolute magnitude assuming an
extinction of $A_V=4.0$ mag. See the text for details.\label{fig:LLL}}
\end{figure}

Figure \ref{fig:LLL} shows the LLL values as a function of age for the
different filters used in this paper. These luminosities were obtained
by identifying the most luminous star on each isochrone for the
relevant passband.  The dark-grey area corresponds to the
`source-dominated' clusters (cf. Fig. 3), assuming a distance $D =
3.9$ Mpc to M82. The light-grey area shows the cluster's maximum
absolute luminosity, assuming an extinction of $A_V=4.0$ mag (roughly
matching the mean extinction value of Melo et al. 2005; see also Mayya
et al. 2008). The upper luminosity limit has been corrected for
extinction using $A_\mathrm{F435W} = 1.317 A_V, A_\mathrm{F555W} =
0.98 A_V$, and $A_\mathrm{F814W} = 0.599 A_V$, i.e., the corresponding
corrections for a G2V-type star using a Cardelli et al. (1989)
extinction law\footnote{see {\tt http://stev.oapd.inaf.it/cmd}.} and a
total-to-selective extinction ratio, $R_V=3.1$.

We see that most of our clusters lie below the LLL, except in the age
range around 2.5--6 Myr. When an extinction correction of $A_V=4$ mag
is used, most clusters still lie below the LLL, particularly in the
$I$ band. This means that, in general, none of our clusters can host
the most luminous star that would be present theoretically for the
given cluster age. This does not mean, however, that our sample
clusters cannot host the occasional massive, evolved, red star (see
for examples, e.g., Bruhweiler et al. 2003; Jamet et al. 2004;
Pellerin 2006). However, in this case, gaps in the sampling of the
cluster mass function are an {\it a posteriori} consequence of
applying the LLL test.

\subsection{Age estimates of incompletely sampled clusters}

In the previous section we established that our clusters are affected
by strong IMF sampling effects, so that age and mass estimates based
on direct comparisons with SSP mean values are not valid (although we
will show below what one would conclude in terms of cluster ages and
the implications for the cluster-formation history in the centre of
M82 if we were to ignore this caveat). As an example, red cluster
colours could be explained as being caused by the absence of massive
stars, but this absence can -- in turn -- be explained by either an
old population in which the massive stars have faded, or a young
population without massive stars (at least not to the extent expected
{\it theoretically}), because of sampling effects (i.e., there are not
enough stars to fully populate the massive-star tail of the IMF). In
an analogous manner, cluster masses cannot be obtained properly, since
the mass-luminosity relation depends on both the age and the cluster
mass itself, and both are unknown quantities.

However, we can still obtain partial information about the clusters
from their integrated luminosities, taking advantage of the
sensitivity of different observables with respect to both stellar
populations and individual stars. In doing so, one needs to take into
account several basic rules, which we will now discuss.

\subsubsection{General considerations and rules}

First, the position of single stars in a colour-colour diagram defines
a partial envelope of all possible conditions. As shown by Cervi\~no
\& Luridiana (2006), the distribution of any possible integrated
luminosity of any cluster containing $N$ stars is the result of $N$
consecutive convolutions of the stellar luminosity function (in units
of flux). Hence, the possible colour distribution should -- eventually
-- converge to the mean value predicted by the SSP models.

Secondly, there is a smooth transition between the positions of single
stars in a colour-colour diagram and the colours of SSPs, and clusters
that are more affected by sampling effects at a given age would cover
a larger region in colour-colour space than clusters containing more
stars (which are, hence, less affected by IMF sampling statistics). In
practice, this situation can be visualised as a gradual collapse of
the possible cluster colours towards the mean SSP prediction.

Thirdly, the colour-colour diagram used should show an asymmetric
trend, in the sense that young clusters and hot stars have bluer
colours than older clusters and cool stars and, in general, the bluest
stars in a cluster correspond to the main-sequence turn off (for
simplicity, we choose to ignore issues related to the nebular
continuum in photo-ionised clusters here; the main effect of nebular
flux would be to lead to redder cluster colours in a similar fashion
as reddening due to extinction. As a consequence, some perceived
`intermediate'-age clusters may therefore be young).

\begin{figure}
\includegraphics[width=\columnwidth]{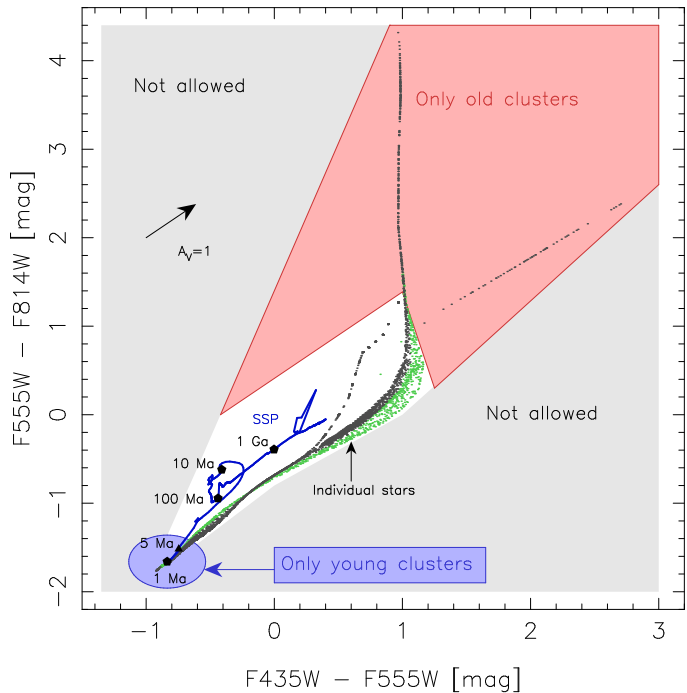}
\caption{Graphical representation of the boundary conditions. The blue
track shows the evolution of well-populated SSP models (from the
bottom-left corner of the figure to redder colours as a function of
increasing age). The black pentagrams highlight SSPs of ages as
indicated. The red area defines the region within which {\it only} old
($\ge 10^8$ yr) clusters can reside; similarly, the blue area
encompasses the region where we expect only young ($\le 5$ Myr-old)
clusters. The black dots show the positions of extremely
underpopulated clusters (i.e., `clusters' composed of single stars,
based on the isochrones discussed in Section \ref{models.sect}). The
green dots represent stars with luminosities below those of the
clusters in our sample, whereas the grey dots are stars that are more
luminous than our clusters (in any filter). The grey-shaded areas
cannot host any clusters, since there is no possible combination of
single stars that can produce such cluster colours (see the text for
details). The extinction vector for $A_V=1$ mag is shown for
reference.}
\label{generalcons.fig}
\end{figure}

These general considerations are shown graphically in
Fig. \ref{generalcons.fig}.

Since we are in a situation where not all stars defining the isochrone
may be present in the cluster, we are bound by a number of additional
rules. In essence, we need to work with the general considerations
outlined above, but we also need to take into account only those stars
that would be present in the cluster. This leads to two basic rules.

\begin{enumerate}
\item Only individual stars {\it less} luminous than the cluster may
be considered in order to define the region in colour-colour space
where a cluster of a given age could reside.
\item No cluster can have colours that are not covered by either the
individual stars or the SSP models.
\end{enumerate}

\subsection{Application of the LLL test}

\begin{figure}
\includegraphics[width=\columnwidth]{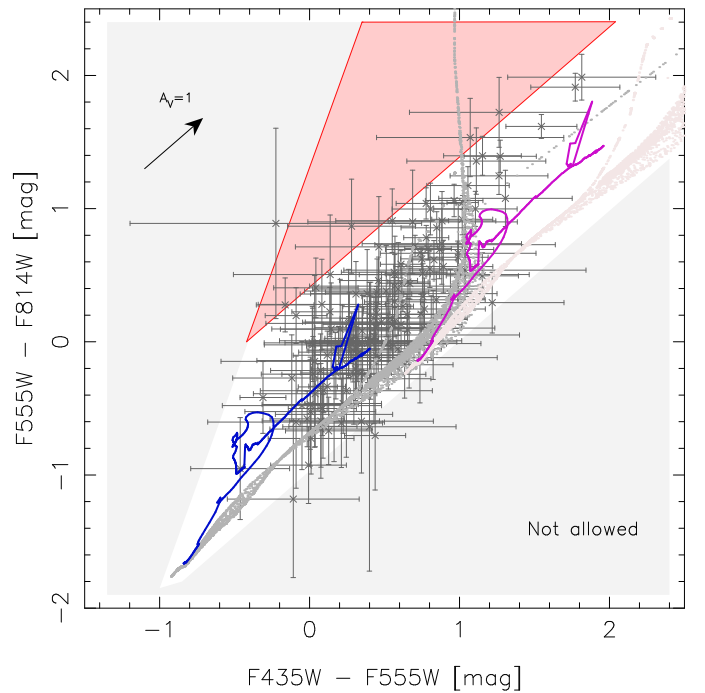}
\caption{Observed colour-colour diagram for the nuclear clusters in
M82. The blue and purple tracks show the evolution of SSP models
affected by $A_V = 0$ and 4 mag, respectively; the dark and light-grey
dots represent the colours of individual stars affected by similar
extinction values. The red shaded area is the region where we would
expect only older ($\ge 10^8$ yr) clusters to be found; the extinction
vector for $A_V = 1$ mag is shown for reference.}
\label{obscolours.fig}
\end{figure}

In Fig. \ref{obscolours.fig} we show the colour-colour diagram of the
observed clusters in the M82 nucleus. We have also included the
positions of the SSP models and of individual stars for a foreground
extinction of $A_V = 0$ and 4 mag, respectively. For $A_V = 0$ mag, it
appears that most of the clusters can be explained by an old age ($\ga
10^8$ yr). However, if we take into account the limitation imposed by
the cluster luminosities, none of these clusters can host stars more
massive than 6 M$_\odot$, and the extreme $(B-V)$ and $(V-I)$ values
are $\sim$1 mag. In this situation, neither clusters redder than
$(B-V) = 1$ mag, nor those redder than the $(V-I)$ values defined by
the SSP models can be explained. This implies that at least some
amount of extinction is required to match the distribution of the
clusters in colour-colour space.

For $A_V = 4$ mag, the model curve seems to imply that all clusters
are young ($\le 5$ Myr old). However, this choice of extinction value
cannot explain any cluster with $(B-V) < 0.5$ mag, for which less
extinction would be required.

Whatever the case, clusters that are redder in $(V-I)$ than the
envelope of the SSP models with variable extinction (i.e., those
redder than a line with a slope $\ga 1$), are most likely explained by
the presence of individual stars in the $(V-I)$ tail at $(B-V) = 1$
mag for $A_V = 0$ mag. This tail corresponds to stars older than 100
Myr and, hence, these are `old' clusters. In this case, a moderate
amount of extinction is needed if the clusters are to host such
luminous red stars. However, we note that in view of the significant
photometric uncertainties, some of the observed redder clusters could
be consistent with younger clusters ($\sim 10^7$--$10^8$ yr old)
affected by varying amounts of extinction.

\subsection{Application of standard SSP analysis}

Although we are clearly aware of the IMF sampling issues discussed in
the previous section, we will now apply the `standard' SSP analysis to
our cluster photometry, in order to show what effects ignoring IMF
sampling effects would have on the derived results. In other words, we
will simply obtain the cluster ages (and their masses) assuming that
their IMFs are well populated, without resorting to the LLL test
discussed above. This will give us a good handle on the age
uncertainties introduced by ignoring IMF sampling effects, which we
will highlight where appropriate and relevant.

\subsection{Age and mass distributions}
\label{ages.sect}

In a series of recent papers, we developed a sophisticated tool for
star cluster analysis based on broad-band SEDs, {\sc AnalySED}, which
we tested extensively both internally (de Grijs et al. 2003a,b; Anders
et al. 2004b) and externally (de Grijs et al. 2005), using both
theoretical and observed young to intermediate-age ($\lesssim 3 \times
10^9$ yr) star cluster SEDs, and the {\sc galev} SSP models (Kurth et
al. 1999; Schulz et al. 2002). The accuracy has been further increased
for younger ages by the inclusion of an extensive set of nebular
emission lines, as well as gaseous continuum emission (Anders \&
Fritze-von Alvensleben 2003). We concluded that the {\it relative}
ages and masses within a given cluster system can be determined to a
very high accuracy, depending on the specific combination of passbands
used (Anders et al. 2004b). Even when comparing the results of
different groups using the same data set, we can retrieve any
prominent features in the cluster age and mass distributions to within
$\Delta \langle \log( {\rm Age / yr} ) \rangle \le 0.35$ and $\Delta
\langle \log( M_{\rm cl} / {\rm M}_\odot ) \rangle \le 0.14$,
respectively (de Grijs et al. 2005), which confirms that we understand
the uncertainties associated with the use of our {\sc AnalySED} tool
to a very high degree -- provided that our sample clusters have
well-populated IMFs.

We therefore applied the {\sc AnalySED} approach to our full set of
broad-band $UBVI$ cluster SEDs (we note that $U$-band photometry is
only available for a subset of 127 clusters), assuming a Kroupa (2001)
IMF (with a low-mass cut-off at 0.1 M$_\odot$) and Padova
isochrones. Four-passband photometry is, in general, sufficient to
yield robust cluster parameters with reasonable uncertainties (Anders
et al. 2004b). However, given the highly variable galactic background
in our field of view (see Fig. \ref{m82core.fig}), we decided to fix
the model cluster metallicities to the solar value, leaving both the
cluster ages and extinction values as free parameters. The shape of
the broad-band SEDs constrains the ages and extinction values, whereas
the absolute flux level results in the corresponding cluster
masses. Despite the significant photometric uncertainties, generally
caused by the highly variable background, our SED matching approach
was fairly successful in converging to a reasonably well-determined
set of ages: of the 152 cluster candidates selected, we obtained ages
for 142.

\begin{figure}
\includegraphics[width=\columnwidth]{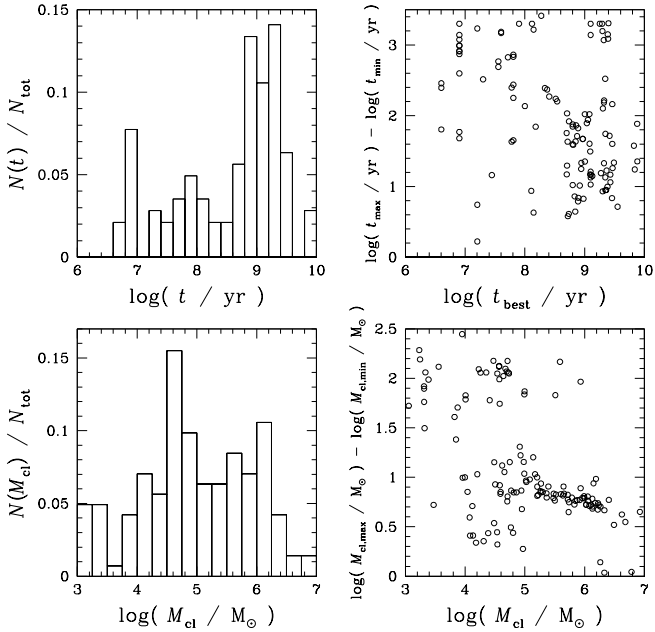}
\caption{{\it (top left and bottom left)} Age and mass distributions
of the clusters in the M82 starburst core, based on the
straightforward application of SED matching using standard SSP models;
{\it (top right and bottom right:} Full ($1\sigma$) uncertainty ranges
in age and mass, as a function of cluster age and mass, respectively.
See the text for details.}
\label{ages.fig}
\end{figure}

The resulting histogram of the best-fitting cluster ages is shown in
the top left-hand panel of Fig. \ref{ages.fig}. The cluster age
distribution shows a peak around log( Age / yr ) $\sim 9$, with a
broad distribution (of order 40\% of the sample clusters) extending
towards younger ages. Whether or not the apparent peak(s) at younger
ages is (are) real depends on the uncertainties in our age
estimates.\footnote{It is well known that broad-band SED fitting
results in artefacts in the cluster age distribution. This is
predominantly caused by specific features in the SSP models, such as
the onset and presence of red giant branch or asymptotic giant branch
stars at, respectively, $\sim 10$ and $\sim 100$ Myr (e.g., Bastian et
al. 2005). In addition, the peak in Fig. \ref{ages.fig} at an age of
$\sim$10 Myr may be an artefact caused by the rapid evolution of
stellar populations around $\sim$10--30 Myr (e.g., Lee et al. 2005).}
The {\sc AnalySED} output also provides realistic ($1\sigma$)
uncertainty estimates (see Anders et al. 2004b). In the top right-hand
panel of Fig. \ref{ages.fig} we visualise the full uncertainty ranges
for all our clusters, as a function of age. We emphasise that these
uncertainties represent the full range from the minimum to the maximum
ages allowed by the fitting routines; because of the logarithmic
representation, we note that the uncertainty ranges are asymmetrical,
however, so that it is impractical to show average values. As an
aside, we also note that our results are not dominated by the
age-extinction degeneracy inherent to the use of broad-band
photometry, given that there is no discernible trend between the
resulting cluster ages and their extinction values (a trend would be
expected if this degeneracy were important). The use of both the
$U$-band filter and a red optical passband ($I$) minimises any
residual age-extinction degeneracy (e.g., Anders et al. 2004b).

We will now use these uncertainty estimates to provide an independent
cross-check on the accuracy of our age determinations. Cluster A1 was
found to have an age of $5.2 \pm 1.5$ Myr (and $E(B-V) =
0.85^{+0.05}_{-0.65}$ mag), which is -- given the uncertainties -- in
good agreement with the age estimate of $6.4 \pm 0.5$ Myr of Smith et
al (2006), based on spectroscopy and for an extinction of $E(B-V) =
1.35 \pm 0.15$ mag. For cluster B2-1 (de Grijs et al. 2001; Smith et
al. 2006), the age derived here ($2.0 \pm 1.0$ Gyr; for $E(B-V) =
0.60^{+0.10}_{-0.20}$ mag) is also consistent with previous age
determinations: Smith et al. (2006) derived $\log(\mbox{Age yr}^{-1})
= 8.54 \pm 0.8$ for $E(B-V)=0.97 \pm 0.51$ mag (see also Smith et
al. 2007), while de Grijs et al. (2001, 2003c) obtained
$\log(\mbox{Age yr}^{-1}) = 9.7$--9.76 based on a small to negligible
amount of extinction.

For completeness, the bottom left-hand panel of Fig. \ref{ages.fig}
shows the corresponding mass distribution of the nuclear M82 clusters
in our sample. This shows that there is a broad distribution of
cluster masses around $10^5$ M$_\odot$, not too dissimilar to the
cluster mass distribution in the post-starburst region `B' some
0.5--1.0 kpc from the centre of M82 (e.g., de Grijs et al. 2001,
2003c). Although there is a non-negligible fraction of clusters more
massive than $\sim 10^6$ M$_\odot$ (which may, hence, qualify as
SSCs), there is also a significant complement of fairly low-mass
objects, down to $\sim 10^3$ M$_\odot$. The bottom right-hand panel
shows the full mass ranges obtained for our sample clusters from the
{\sc AnalySED} fits, in a similar visualisation as shown in the top
right-hand panel for the age uncertainties. A quick comparison of the
general differences between both right-hand panels shows that, indeed,
the average uncertainties in the cluster masses are relatively smaller
than those for the ages, at least for masses $\ga 10^5$ M$_\odot$ (in
support of de Grijs et al. 2005).

\subsection{Constraints from H$\alpha$ imaging}

We are fortunate in also having obtained an H$\alpha$ image of the
galaxy's starburst core. The addition of H$\alpha$ photometry allows
us to determine the number of sample clusters showing an H$\alpha$
excess, which in turn will help us to constrain the cluster ages to
$\la 8$ Myr. Of the 152 clusters in our sample, nineteen show a clear
H$\alpha$ excess exceeding the 1$\sigma$ photometric uncertainties of
the clusters' total H$\alpha$ flux. This relatively small number does
not necessarily imply that most objects are faint in H$\alpha$ flux,
but rather that the photometric uncertainties are generally
significant due to the highly variable background (see
Fig. \ref{m82core.fig}). For those clusters with a statistically
significant H$\alpha$ excess, the corresponding ages based on their
broad-band SEDs are in all cases consistent with them being younger
than $\sim 10$ Myr.

Of these 19 clusters with a significant H$\alpha$ excess, twelve are
located in or near region C. This is consistent with the observation
that region C might be the youngest part of the starburst (Section
2.2), although our age resolution does not allow us to conclude this
robustly; for all practical purposes, we may assume that regions A and
C are of similar (young) age (see also O'Connell \& Mangano 1978;
O'Connell et al. 1995). Alternatively, region C may represent the
actual core region generating the minor-axis H$\alpha$ `plume'
coincident with the minor-axis `superwind' (cf. McCarthy et al. 1987;
O'Connell et al. 1995; Westmoquette et al. 2007b). The right-hand
panel of Fig. \ref{sources.fig} shows the distribution of the
H$\alpha$ excess across the entire region. In this representation,
region C clearly stands out in H$\alpha$ emission, so it may indeed be
the youngest core region.

\section{Summary and Discussion}

We have presented high-resolution {\sl HST}/ACS imaging in four
filters (Johnson-Cousins equivalent $B, V$, and $I$ broad bands, and
an H$\alpha$ narrow-band filter), as well as subsequently acquired
$U$-band images, of the nuclear starburst region of M82. The high
spatial resolution of the {\sl HST} images has allowed us to explore
the central area of this galaxy in unparalleled detail. We find a
complex system of $\sim$150 star clusters in the inner few 100 pc of
M82, encompassing sections A, C, D, and E (O'Connell \& Mangano
1978). At face value, the resulting cluster age distribution gives us
a strong handle on the global star cluster formation intensity across
the galaxy's core region, which can then be compared to the field-star
formation history.

As the archetypal local starburst galaxy, M82 is believed to have
undergone a tidal interaction with its neighbour, M81, $\sim 10^8$ yr
ago (e.g., Gottesman \& Weliachew 1977; O'Connell \& Mangano 1978; Lo
et al. 1987; Yun et al. 1993, 1994), providing the means of activation
of the intense episodes of starburst activity witnessed in M82. It has
been suggested that this interaction led to the ISM of M82
experiencing large-scale torques, coupled with a loss of angular
momentum as it was transferred towards the dynamical centre of the
galaxy (FS03; also based on numerical simulations: e.g., Sundelius et
al. 1987; Noguchi 1987, 1988; Mihos \& Hernquist 1996). This led to an
increased cloud-cloud collision rate in the disk of the galaxy, and
large amounts of material accumulating and undergoing compression in
the innermost regions, perhaps resulting in the first starburst
episode (FS03). 

We find (tentative) evidence of an enhanced cluster-formation epoch
associated with the first starburst event (although we caution that
the age uncertainties are significant), believed to have occurred
about 100 Myr ago (or possibly as recently as $\sim 30$ Myr ago; Rieke
et al. 1993). We do not find any strong evidence\footnote{The apparent
peak in Fig. \ref{ages.fig} at an age of $\sim$10 Myr is most likely
due to the residual effects of (i) the age-extinction degeneracy
(which is particularly important at these young ages), and (ii) the
fact that the youngest isochrone included in the {\sc AnalySED} tool
is of an age of $2.5 \times 10^6$ yr, with limited age resolution at
these ages. In addition, we remind the reader that we have adopted a
fixed, solar metallicity, which may also give rise to a small
age-metallicity degeneracy (in view of the variability of the galactic
background and the associated photometric uncertainties, it is
impractical to introduce an additional free parameter in the fits; the
metallicity is expected to be least variable of our choice of possible
free parameters). The consequence of these effects is that the
youngest clusters tend to accumulate close to the lower age cut-off,
and hence this peak should be taken with extreme caution (see, e.g.,
de Grijs et al. 2003a for a detailed analysis).} of enhanced star
cluster formation at an age of $\sim 5$ Myr, however, believed to be
the epoch when the more recent starburst event transpired in the
region (FS03; supported by Smith et al. 2006). Perhaps this is due to
the relatively small region studied here, in that it could be possible
that the clusters analysed are not spatially associated with the more
recent starburst event.

F\"orster Schreiber et al. (2003) indeed suggest that the two events
took place in different regions, with the first happening in the
centre of M82 and the second occurring predominantly in a
circumnuclear ring and along the stellar bar. Alternatively, it could
be the case that cluster formation is not always coincident with
enhanced star-formation episodes. This type of scenario is, in fact,
not unreasonable to expect, as evidenced by, e.g., the observed
disparities between the cluster and field-star age distributions in
the Magellanic Clouds and in NGC 1569 (see Anders et al. 2004a for a
discussion regarding the latter galaxy). In particular, the Large
Magellanic Cloud exhibits a well-known gap in the cluster age
distribution, yet the age distribution of the field stellar population
appears more continuous; the field-star and cluster formation
histories are clearly very different (e.g., Olszewski et al. 1996;
Geha et al. 1998; Sarajedini 1998; and references therein). Although
the case is less clear-cut for the Small Magellanic Cloud, Rafelski \&
Zaritsky (2005) provide tentative evidence that the cluster and
field-star age distributions are also significantly different in this
system (see also Gieles et al. 2007).

We caution, however, that our cluster age estimates may be severely
affected by IMF stochasticity, given that the luminosities of most of
our sample clusters imply that their mass functions are not fully
sampled up to the most luminous stars included in the theoretical
stellar isochrones. We show the effects of statistical IMF sampling
issues in colour-colour space. This should be taken as a strong
warning to anyone (including ourselves) attempting to apply standard
SSP analysis to integrated cluster photometry of any but the most
massive star clusters. As a consequence, the age estimates derived
based on standard SSP analysis should be taken with extreme caution.
It is most likely that our straightforward SSP-based age
determinations are affected by (i) residual effects due to the
age-extinction degeneracy, despite the availability of $U$-band and
H$\alpha$ observations, and -- more importantly -- (ii) significant
stochastic effects.

It is our intention that this paper be taken by the community as a
strong warning to consider stochasticity in the IMF for young clusters
with undersampled stellar mass functions much more seriously than has
been done to date (including by ourselves; e.g., de Grijs et
al. 2003a,b).

\section*{Acknowledgements}

We thank Peter Anders for his help with the analysis of our broad-band
cluster photometry. SB thanks Sarah Moll for her help using the {\sc
ishape} function of {\sc BAOlab}. SB and RdG thank Paul Kerry for his
invaluable help with all sorts of software-related issues. RdG
acknowledges useful discussions with Simon Goodwin, and hospitality at
and research support from the International Space Science Institute in
Bern (Switzerland). MC is supported by the Spanish MCyT and by FEDER
funding of project AYA2007-64712, and by a Ram\'on y Cajal
fellowship. The {\sc cmd2.0} tool was developed by Leo Girardi. We are
grateful to the anonymous referee for a careful and constructive
report that led to significant improvements in both our interpretation
and the presentation of our results. This paper is based on archival
observations with the NASA/ESA {\sl Hubble Space Telescope}, obtained
at the Space Telescope Science Institute (STScI), which is operated by
the Association of Universities for Research in Astronomy,
Inc. (AURA), under NASA contract NAS5-26555. This research project was
funded by the Nuffield Foundation through Undergraduate Summer Bursary
URB/34412. This research has made use of NASA's Astrophysics Data
System Abstract Service.


\begin{thebibliography}{}

\bibitem[\protect\citeauthoryear{Anders \&
Fritze-von~Alvensleben}{2003}]{2003A&A...401.1063A} Anders, P., \&
Fritze-von~Alvensleben, U. 2003, A\&A, 401, 1063

\bibitem[\protect\citeauthoryear{Anders et
al.}{2004}]{2004MNRAS.347...17A} Anders, P., de Grijs, R.,
Fritze-von~Alvensleben, U., \& Bissantz, N. 2004a, MNRAS, 347, 17

\bibitem[\protect\citeauthoryear{Anders et
al.}{2004}]{2004MNRAS.347..196A} Anders, P., Bissantz, N.,
Fritze-von~Alvensleben, U., \& de Grijs, R. 2004b, MNRAS, 347, 196

\bibitem[\protect\citeauthoryear{Anders, Gieles, \& de
Grijs}{2006}]{2006A&A...451..375A} Anders, P., Gieles, M., \& de
Grijs, R. 2006, A\&A, 451, 375

\bibitem[\protect\citeauthoryear{Arp \&
Sandage}{1985}]{1985AJ.....90.1163A} Arp, H., \& Sandage, A. 1985, AJ,
90, 1163

\bibitem[]{} Bastian N., Gieles M., Lamers H.J.G.L.M., Scheepmaker
R.A., de Grijs R., 2005, A\&A, 431, 905

\bibitem[\protect\citeauthoryear{Bertelli et
al.}{1994}]{1994A&AS..106..275B} Bertelli, G., Bressan, A., Chiosi,
C., Fagotto, F., \& Nasi, E. 1994, A\&AS, 106, 275

\bibitem[\protect\citeauthoryear{Brouillet et
al.}{1991}]{1991A&A...242...35B} Brouillet, N., Baudry, A., Combes,
F., Kaufman, M., \& Bash, F. 1991, A\&A, 242, 35

\bibitem[\protect\citeauthoryear{Bruhweiler, Miskey, \& Smith
Neubig}{2003}]{2003AJ....125.3082B} Bruhweiler, F.~C., Miskey, C.~L.,
\& Smith Neubig M. 2003, AJ, 125, 3082

\bibitem[\protect\citeauthoryear{Cardelli, Clayton, \&
Mathis}{1989}]{Cardelli89} Cardelli, J.~A., Clayton, G.~C., \& Mathis,
J.~S. 1989, ApJ, 345, 245

\bibitem[\protect\citeauthoryear{Carlson et
al.}{1998}]{1998AJ....115.1778C} Carlson, M.~N., Holtzman, J. A.,
Watson, A. M., et al. 1998, AJ, 115, 1778

\bibitem[\protect\citeauthoryear{Cervi{\~n}o, Luridiana, \&
Castander}{2000}]{CLC00} Cervi{\~n}o, M., Luridiana, V., \& Castander,
F.~J. 2000, A\&A, 360, L5

\bibitem[Cervi{\~ n}o et al.(2001)]{2001AAp...376...422C} Cervi{\~n}o,
M., G{\' o}mez-Flechoso, M.~A., Castander, F.~J., et al. 2001 A\&A,
376, 422

\bibitem[\protect\citeauthoryear{Cervi{\~n}o et al.}{2002}]{CVGLMH02}
Cervi{\~n}o, M., Valls-Gabaud, D., Luridiana, V., \& Mas-Hesse,
J.~M. 2002, A\&A, 381, 51

\bibitem[\protect\citeauthoryear{Cervi{\~n}o \&
Luridiana}{2004}]{CL04} Cervi{\~n}o, M., \& Luridiana, V. 2004, A\&A,
413, 145

\bibitem [\protect\citeauthoryear{Cervi\~{n}o \&
Luridiana}{2006}]{CL06} Cervi\~{n}o, M., \& Luridiana, V. 2006, A\&A,
451, 475

\bibitem[\protect\citeauthoryear{Cervi{\~n}o \&
Valls-Gabaud}{2008}]{CVG08} Cervi{\~n}o, M. Valls-Gabaud, D. 2008, in:
Young massive star clusters -- Initial conditions and environments,
eds.  E. P\'erez, R. de Grijs, \& R. M. Gonz\'alez Delgado, Springer:
Dordrecht, in press (arXiv:0802.3213v1)

\bibitem[\protect\citeauthoryear{Conti, Leitherer, \&
Vacca}{1996}]{1996ApJ...461L..87C} Conti, P.~S., Leitherer, C., \&
Vacca, W.~D. 1996, ApJ, 461, L87

\bibitem[\protect\citeauthoryear{de Grijs et
al.}{2000}]{2000AJ....119..681D} de Grijs, R., O'Connell, R.~W.,
Becker, G.~D., Chevalier, R.~A., \& Gallagher, J.~S., {\sc iii} 2000,
AJ, 119, 681

\bibitem[\protect\citeauthoryear{de Grijs, O'Connell, \&
Gallagher}{2001}]{2001AJ....121..768D} de Grijs, R., O'Connell, R.~W.,
\& Gallagher, J.~S., {\sc iii} 2001, AJ, 121, 768

\bibitem[\protect\citeauthoryear{de Grijs et
al.}{2003}]{2003MNRAS.343.1285D} de Grijs, R., Anders, P., Bastian,
N., et al. 2003a, MNRAS, 343, 1285

\bibitem[\protect\citeauthoryear{de Grijs et
al.}{2003}]{2003MNRAS.342..259D} de Grijs, R., Fritze-von~Alvensleben,
U., Anders, P., et al. 2003b, MNRAS, 342, 259

\bibitem[\protect\citeauthoryear{de Grijs, Bastian, \& 
Lamers}{2003}]{2003ApJ...583L..17D} de Grijs, R., Bastian, N., \& Lamers, 
H.~J.~G.~L.~M. 2003c, ApJ, 583, L17 

\bibitem[\protect\citeauthoryear{de Grijs et
al.}{2005}]{2005MNRAS.359..874D} de Grijs, R., Anders, P., Lamers,
H.~J.~G.~L.~M., et al. 2005, MNRAS, 359, 874

\bibitem[\protect\citeauthoryear{de Grijs \&
Parmentier}{2007}]{2007ChJAA...7..155D} de Grijs, R., \& Parmentier,
G. 2007, ChJAA, 7, 155

\bibitem[\protect\citeauthoryear{Fabbiano \&
Trinchieri}{1984}]{1984ApJ...286..491F} Fabbiano, G., \& Trinchieri,
G. 1984, ApJ, 286, 491

\bibitem[\protect\citeauthoryear{F{\"o}rster Schreiber et
al.}{2003}]{2003ApJ...599..193F} F{\"o}rster Schreiber, N.~M., Genzel,
R., Lutz, D., \& Sternberg, A. 2003, ApJ, 599, 193

\bibitem[\protect\citeauthoryear{Fritze-von~Alvensleben \&
Gerhard}{1994}]{1994A&A...285..775F} Fritze-von~Alvensleben, U., \&
Gerhard, O.~E. 1994, A\&A, 285, 775

\bibitem[\protect\citeauthoryear{Gallagher \&
Smith}{1999}]{1999MNRAS.304..540G} Gallagher, J.~S., {\sc iii}, \&
Smith, L.~J. 1999, MNRAS, 304, 540

\bibitem[\protect\citeauthoryear{Geha et
al.}{1998}]{1998AJ....115.1045G} Geha, M.~C., Holtzman, J. A., Mould,
J. R., et al. 1998, AJ, 115, 1045

\bibitem[\protect\citeauthoryear{Gieles, Lamers, \& Portegies
Zwart}{2007}]{2007ApJ...668..268G} Gieles, M., Lamers, H.~J.~G.~L.~M.,
\& Portegies Zwart, S.~F. 2007, ApJ, 668, 268

\bibitem[\protect\citeauthoryear{Girardi et
al.}{2000}]{2000A&AS..141..371G} Girardi, L., Bressan, A., Bertelli,
G., \& Chiosi, C. 2000, A\&AS, 141, 371

\bibitem[\protect\citeauthoryear{Girardi et
al.}{2002}]{2002A&A...391..195G} Girardi, L., Bertelli, G., Bressan,
A., et al. 2002, A\&A, 391, 195

\bibitem[\protect\citeauthoryear{Gottesman \&
Weliachew}{1977}]{1977ApJ...211...47G} Gottesman, S.~T., \& Weliachew,
L. 1977, ApJ, 211, 47

\bibitem[\protect\citeauthoryear{Ho \&
Filippenko}{1996}]{1996ApJ...472..600H} Ho, L.~C., \& Filippenko,
A.~V. 1996, ApJ, 472, 600

\bibitem[\protect\citeauthoryear{Ho}{1997}]{1997RMxAC...6....5H} Ho,
L.~C. 1997, RMxAC, 6, 5

\bibitem[\protect\citeauthoryear{Holtzman et
al.}{1992}]{1992AJ....103..691H} Holtzman, J.~A., Faber, S. M., Shaya,
E. J., et al. 1992, AJ, 103, 691

\bibitem[\protect\citeauthoryear{Hunter, O'Connell, \&
Gallagher}{1994}]{1994AJ....108...84H} Hunter, D.~A., O'Connell,
R.~W., \& Gallagher, J.~S., {\sc iii} 1994, AJ, 108, 84

\bibitem[\protect\citeauthoryear{Hunter et
al.}{2000}]{2000AJ....120.2383H} Hunter, D.~A., O'Connell, R.~W.,
Gallagher, J.~S., {\sc iii}, \& Smecker-Hane, T.~A. 2000, AJ, 120,
2383

\bibitem[\protect\citeauthoryear{Jamet et
al.}{2004}]{2004A&A...426..399J} Jamet, L., P{\'e}rez, E.,
Cervi{\~n}o, M., Stasi{\'n}ska, G., Gonz{\'a}lez Delgado, R.~M., \&
V{\'{\i}}lchez, J.~M. 2004, A\&A, 426, 399

\bibitem[\protect\citeauthoryear{Jansen et
al.}{1994}]{1994MNRAS.270..373J} Jansen, R.~A., Knapen, J.~H.,
Beckman, J.~E., Peletier, R.~F., \& Hes, R. 1994, MNRAS, 270, 373

\bibitem[\protect\citeauthoryear{Keto, Ho, \&
Lo}{2005}]{2005ApJ...635.1062K} Keto, E., Ho, L.~C., \& Lo,
K.-Y. 2005, ApJ, 635, 1062

\bibitem[Krist \& Hook (1997)]{krist97} Krist, J., \& Hook, R. 1997,
The Tiny Tim User's Guide, Baltimore: STScI

\bibitem[\protect\citeauthoryear{Kroupa}{2001}]{kroupa} Kroupa, P.
2001, MNRAS, 322, 231

\bibitem[]{} Kurth, O.M., Fritze-von Alvensleben, U., \& Fricke,
K.J. 1999, A\&AS, 138, 19

\bibitem[\protect\citeauthoryear{Kurucz}{1992}]{Kurucz} Kurucz, R.~L.
1992, IAUS, 149, 225

\bibitem[Larsen(1999)]{1999A&AS..139..393L} Larsen. S.~S. 1999, A\&AS,
139, 393

\bibitem[\protect\citeauthoryear{Lee, Chandar, \&
Whitmore}{2005}]{2005AJ....130.2128L} Lee, M.~G., Chandar, R., \&
Whitmore, B.~C. 2005, AJ, 130, 2128

\bibitem[\protect\citeauthoryear{Lo et
al.}{1987}]{1987ApJ...312..574L} Lo, K.~Y., Cheung, K.~W., Masson,
C.~R., et al. 1987, ApJ, 312, 574

\bibitem[\protect\citeauthoryear{Lynds \&
Sandage}{1963}]{1963ApJ...137.1005L} Lynds, C.~R., \& Sandage,
A.~R. 1963, ApJ, 137, 1005

\bibitem[\protect\citeauthoryear{Ma et
al.}{2006}]{2006MNRAS.368.1443M} Ma, J., de Grijs, R., Yang, Y., et
al. 2006, MNRAS, 368, 1443

\bibitem[\protect\citeauthoryear{Marigo et al.}{2007}]{Maetal07}
Marigo, P., Girardi, L., Bressan, A., et al. 2008, A\&A, in press
(arXiv:0711.4922v1)

\bibitem[\protect\citeauthoryear{Mayya et al.}{2008}]{Mayya08} Mayya,
Y. D., Romano, R., Rodr\'\i guez-Merino, L. H., et al. 2008, ApJ, in
press (arXiv:0802.1922v1)

\bibitem[\protect\citeauthoryear{McCarthy, van Breugel, \&
Heckman}{1987}]{1987AJ.....93..264M} McCarthy, P.~J., van Breugel, W.,
\& Heckman, T. 1987, AJ, 93, 264

\bibitem[]{} McCrady, N., \& Graham, J. R 2007, ApJ, 663, 844

\bibitem[\protect\citeauthoryear{McKeith et
al.}{1995}]{1995A&A...293..703M} McKeith, C.~D., Greve, A., Downes,
D., \& Prada, F. 1995, A\&A, 293, 703

\bibitem[\protect\citeauthoryear{McLeod et
al.}{1993}]{1993ApJ...412..111M} McLeod, K.~K., Rieke, G.~H., Rieke,
M.~J., \& Kelly, D.~M. 1993, ApJ, 412, 111

\bibitem[\protect\citeauthoryear{Melnick, Moles, \&
Terlevich}{1985}]{1985A&A...149L..24M} Melnick, J., Moles, M., \&
Terlevich, R.  1985, A\&A, 149, L24

\bibitem[\protect\citeauthoryear{Melo et
al.}{2005}]{2005ApJ...619..270M} Melo, V.~P., Mu{\~n}oz-Tu{\~n}{\'o}n,
C., Ma{\'{\i}}z-Apell{\'a}niz, J., \& Tenorio-Tagle, G. 2005, ApJ,
619, 270

\bibitem[\protect\citeauthoryear{Meurer et
al.}{1992}]{1992AJ....103...60M} Meurer, G.~R., Freeman, K.~C.,
Dopita, M.~A., \& Cacciari, C. 1992, AJ, 103, 60

\bibitem[\protect\citeauthoryear{Mutchler et
al.}{2007}]{2007PASP..119....1M} Mutchler, M., Bond, H. E., Christian,
C. A., et al. 2007, PASP, 119, 1

\bibitem[\protect\citeauthoryear{Mihos \&
Hernquist}{1996}]{1996ApJ...464..641M} Mihos, J.~C., \& Hernquist,
L. 1996, ApJ, 464, 641

\bibitem[\protect\citeauthoryear{Noguchi}{1987}]{1987MNRAS.228..635N} 
Noguchi, M. 1987, MNRAS, 228, 635 

\bibitem[\protect\citeauthoryear{Noguchi}{1988}]{1988A&A...201...37N} 
Noguchi, M. 1988, A\&A, 201, 37 

\bibitem[\protect\citeauthoryear{O'Connell \&
Mangano}{1978}]{1978ApJ...221...62O} O'Connell, R.~W., \& Mangano,
J.~J. 1978, ApJ, 221, 62

\bibitem[\protect\citeauthoryear{O'Connell, Gallagher, \&
Hunter}{1994}]{1994ApJ...433...65O} O'Connell, R.~W., Gallagher,
J.~S., {\sc iii}, \& Hunter, D.~A. 1994, ApJ, 433, 65

\bibitem[\protect\citeauthoryear{O'Connell et
al.}{1995}]{1995ApJ...446L...1O} O'Connell, R.~W., Gallagher, J.~S.,
{\sc iii}, Hunter, D.~A., \& Colley, W.~N. 1995, ApJ, 446, L1

\bibitem[\protect\citeauthoryear{Olszewski, Suntzeff, \&
Mateo}{1996}]{1996ARA&A..34..511O} Olszewski, E.~W., Suntzeff, N.~B.,
\& Mateo, M. 1996, ARA\&A, 34, 511

\bibitem[\protect\citeauthoryear{Pellerin}{2006}]{2006AJ....131..849P}
Pellerin, A. 2006, AJ, 131, 849

\bibitem[\protect\citeauthoryear{Pessev et al.}{2008}]{Pessev08}
Pessev, P. M., Goudfrooij, P., Puzia, T. H., \& Chandar, R. 2008,
MNRAS, in press (arXiv:0801.2375v1)

\bibitem[\protect\citeauthoryear{Rafelski \&
Zaritsky}{2005}]{2005AJ....129.2701R} Rafelski, M., \& Zaritsky,
D. 2005, AJ, 129, 2701

\bibitem[\protect\citeauthoryear{Rieke \&
Lebofsky}{1985}]{1985ApJ...288..618R} Rieke, G.~H., \& Lebofsky,
M.~J. 1985, ApJ, 288, 618

\bibitem[\protect\citeauthoryear{Rieke et
al.}{1980}]{1980ApJ...238...24R} Rieke, G.~H., Lebofsky, M.~J.,
Thompson, R.~I., Low, F.~J., \& Tokunaga, A.~T. 1980, ApJ, 238, 24

\bibitem[\protect\citeauthoryear{Rieke et
al.}{1993}]{1993ApJ...412...99R} Rieke, G.~H., Loken, K., Rieke,
M.~J., \& Tamblyn, P. 1993, ApJ, 412, 99

\bibitem[\protect\citeauthoryear{Sakai \&
Madore}{1999}]{1999ApJ...526..599S} Sakai, S., \& Madore, B.~F. 1999,
ApJ, 526, 599

\bibitem[\protect\citeauthoryear{Sarajedini}{1998}]{1998AJ....116..738S} 
Sarajedini, A. 1998, AJ, 116, 738

\bibitem[\protect\citeauthoryear{Satyapal et
al.}{1995}]{1995ApJ...448..611S} Satyapal, S., Watson, D. M., Pipher,
J. L., et al. 1995, ApJ, 448, 611

\bibitem[\protect\citeauthoryear{Schlegel, Finkbeiner, \&
Davis}{1998}]{1998ApJ...500..525S} Schlegel, D.~J., Finkbeiner, D.~P.,
\& Davis, M. 1998, ApJ, 500, 525

\bibitem[]{} Schulz, J., Fritze-von Alvensleben, U., M\"oller, C.S., \&
Fricke, K.J. 2002, A\&A, 392, 1

\bibitem[\protect\citeauthoryear{Shopbell \&
Bland-Hawthorn}{1998}]{1998ApJ...493..129S} Shopbell, P.~L., \&
Bland-Hawthorn, J. 1998, ApJ, 493, 129

\bibitem[\protect\citeauthoryear{Silich, Tenorio-Tagle, \&
Mu{\~n}oz-Tu{\~n}{\'o}n}{2007}]{2007ApJ...669..952S} Silich, S.,
Tenorio-Tagle, G., \& Mu{\~n}oz-Tu{\~n}{\'o}n, C. 2007, ApJ, 669, 952

\bibitem[\protect\citeauthoryear{Smith et
al.}{2006}]{2006MNRAS.370..513S} Smith, L.~J., Westmoquette, M.~S.,
Gallagher, J.~S., {\sc iii}, et al. 2006, MNRAS, 370, 513

\bibitem[\protect\citeauthoryear{Smith et
al.}{2007}]{2007ApJ...667L.145S} Smith, L.~J., Bastian, N.,
Konstantopoulos, I. S., et al. 2007, ApJ, 667, L145

\bibitem[\protect\citeauthoryear{Strickland, Ponman, \&
Stevens}{1997}]{1997A&A...320..378S} Strickland, D.~K., Ponman, T.~J.,
\& Stevens, I.~R. 1997, A\&A, 320, 378

\bibitem[\protect\citeauthoryear{Sundelius et
al.}{1987}]{1987A&A...174...67S} Sundelius, B., Thomasson, M.,
Valtonen, M.~J., \& Byrd, G.~G. 1987, A\&A, 174, 67

\bibitem[\protect\citeauthoryear{Telesco et
al.}{1991}]{1991ApJ...369..135T} Telesco, C.~M., Joy, M., Dietz, K.,
Decher, R., \& Campins, H. 1991, ApJ, 369, 135

\bibitem[\protect\citeauthoryear{Tenorio-Tagle, Silich, \&
Mu{\~n}oz-Tu{\~n}{\'o}n}{2003}]{2003ApJ...597..279T} Tenorio-Tagle,
G., Silich, S., \& Mu{\~n}oz-Tu{\~n}{\'o}n, C. 2003, ApJ, 597, 279

\bibitem[\protect\citeauthoryear{Watson, Stanger, \&
Griffiths}{1984}]{1984ApJ...286..144W} Watson, M.~G., Stanger, V., \&
Griffiths, R.~E. 1984, ApJ, 286, 144

\bibitem[\protect\citeauthoryear{Watson et
al.}{1996}]{1996AJ....112..534W} Watson, A.~M., Gallagher, J. S., {\sc
iii}, Holtzman, J. A., et al. 1996, AJ, 112, 534

\bibitem[\protect\citeauthoryear{Westmoquette et
al.}{2007}]{2007MNRAS.381..913W} Westmoquette, M.~S., Smith, L.~J.,
Gallagher, J.~S., {\sc iii}, Exter, K.~M. 2007a, MNRAS, 381, 913

\bibitem[\protect\citeauthoryear{Westmoquette et
al.}{2007}]{2007ApJ...671..358W} Westmoquette, M.~S., Smith, L.~J.,
Gallagher, J.~S., {\sc iii}, et al. 2007b, ApJ, 671, 358

\bibitem[\protect\citeauthoryear{Whitmore et
al.}{1993}]{1993AJ....106.1354W} Whitmore, B.~C., Schweizer, F.,
Leitherer, C., Borne, K., \& Robert, C. 1993, AJ, 106, 1354

\bibitem[\protect\citeauthoryear{Whitmore et
al.}{1999}]{1999AJ....118.1551W} Whitmore, B.~C., Zhang, Q.,
Leitherer, C., et al. 1999, AJ, 118, 1551

\bibitem[\protect\citeauthoryear{Yun, Ho, \&
Lo}{1993}]{1993ApJ...411L..17Y} Yun, M.~S., Ho, P.~T.~P., \& Lo,
K.~Y. 1993, ApJ, 411, L17

\bibitem[\protect\citeauthoryear{Yun, Ho, \&
Lo}{1994}]{1994Natur.372..530Y} Yun, M.~S., Ho, P.~T.~P., \& Lo,
K.~Y. 1994, Nature, 372, 530

\end{thebibliography}
\end{document}